In this paper I suggest a possible explanation for the asymmetry of time. In the case that I study, the dynamical laws and the boundary conditions are symmetric, but the behavior of time is not. The underlying mechanism is statistical and closely related to the idea of multiple histories in quantum mechanics, but otherwise rather independent of the particular framework.

Author keywords: multiverse, time's arrow, cosmology, entropy

Je considère une possibilité d'explication pour l'asymétrie temporelle. Dans le cas que j'étudie, la dynamique et les conditions aux limites sont symétriques, mais pas le comportement de la variable temporelle. Le mécanisme sous-jacent est statistique et est étroitement relié à l'idée d'histoires multiples en mécanique quantique, tout en restant assez indépendant du cadre particulier.

# Time's Arrow from the Multiverse Point of View
Martin Tamm

# 1    Introduction

To explain the second law of thermodynamics and the non-symmetric behavior of time is a fundamental problem in contemporary physics. In any situation known to us, the direction towards the future differs from the direction towards the past by the property that in the former entropy increases and in the latter it decreases. The question is: what causes this asymmetric behavior?

In several papers and in particular in his book [1] (see also [2], [3]), Huw Price has analyzed some of the possible answers that physics has come up with, and found that there is still a lot of confusion about what is actually the problem and what should be done to solve it: our very special human perspective on the issue of time tends to make us formulate the questions in the wrong way. In addition, it is not clear where to look for the solution: is it a quantum mechanical or an essentially classical problem? Is it a problem about the boundary conditions of the universe or is it something hidden within the dynamical laws? There may also be a problem with the idea of entropy itself: although we have a very good understanding of this concept in most situations, in cosmology it is not quite clear what it should refer to.

It may very well be that the problem of time's arrow is too complex to have a simple solution in the usual physical or mathematical sense of the word. But it can still be that we can clarify the problem in a convincing way by constructing simple models of the universe where the reason for the break of time-symmetry can be seen clearly.

I have chosen here to work with a bounded model for space-time, even if most votes in recent years have supported open models. The reason is not that I want to argue in favor of a certain model for cosmology at this point. Rather, the motivation is that in this case the underlying global structure of space-time can be considered to be symmetric and finite, which makes the problem with the direction of time come out more clearly.

Many of the ideas in this paper are related to work by other people and have appeared in other places in different form (see Section 2 for some background). In fact, my ambition is not so much to introduce new physical ideas as to launch a different way of thinking about the problem. Although a lot of effort has been made to understand the arrow of time starting from fundamental theories (see e.g. www.fqxi.org for some recent developments), and also a lot of work has been done within statistical mechanics to understand the second law of thermodynamics, somehow there seems to remain a gap between these two methods of attack.

In this paper I will have very little to say about fundamental theories. Rather, the idea is to use general statistical ideas to try to reach another kind of understanding. To illustrate my opinion about what should be done, the following analogy may be appropriate: a central place in the theory of phase transitions is held by the two-dimensional Ising model ([4] and [5]). This is an essentially classical and extremely simplified model of a ferromagnet. Quantum theory comes in only through the quantized magnetic moment, and in a way which fits well with the formalism of classical statistical mechanics. In spite of its classical and very simplified nature, and in spite of the fact that there are much more realistic models for ferromagnetism, no other model has had such an impact on our understanding of the nature of phase transitions. The main reason for this success seems to be that the statistical difficulties of the model happen to be on just the right level of difficulty; it is complicated enough to represent the essential problem, but still simple enough to be within reach for our methods.

A reasonable model for time-asymmetry must by necessity be more complicated than the Ising-model, and this paper is only an attempt to find a kind of starting-point. In fact, what I would like to do is to construct an exact combinatorial model for a small multiverse which could be treated with rigorous mathematical methods. However, so far this task is still too difficult, hence I will in the following rather use statistical arguments. Also, clearly the analogy with statistical mechanics should not be pushed too far. Nevertheless, I will in this paper be concerned with a very simple statistical model for time-asymmetry which is essentially of classical nature. And just like in the Ising model, quantum mechanics enters in a simple way (in this case by assuming multiple histories). In Section 3, I will describe this semi-classical multiverse.

A particular problem with this kind of multiverse is the idea of a state. Although the world is quantum mechanical, and although there is a sound theory for quantum mechanical states, this can be somewhat misleading in the multiverse context, since "parallel universes" are not to be regarded as quantum mechanical states. This is in fact one of the reasons why I have chosen to work with a more primitive classical kind of framework. In addition, I may add that it *is* my belief that (except for the multiverse perspective and perhaps the behavior neat the beginning and the end) the arrow of time can essentially be understood using classical physics.

Although we often speak of *the* arrow of time, there are actually several different arrows, and one of the most important questions is how these are related to each other. In Section 4, I discuss what I consider to be the most important ones.

Another major problem when trying to understand the arrow of time is that it is not quite clear how to measure entropy in a world where the underlying geometry changes, like for instance in an expanding universe. In Section 5 I will therefore discuss the concept of entropy from this point of view. An additional problem is that to make any kind of computations with entropy in cosmology, by necessity we must make enormous simplifications, and the approach in this paper is no exception.

Up to this point, my perspective is in a certain sense traditional, since a direction of time is more or less taken for granted. In Section 6 however, the goal is to present a model which explains time asymmetry starting from completely symmetric laws and boundary conditions. The underlying idea here, as well as in many other papers starting from Boltzmann (see [6]), is to perceive the second law as a consequence of the fact that the universe, as time evolves, passes from un-probable states to more probable ones. However, formulated in this way, the idea already has a definite idea of time direction built in to it, which tends to make it worthless in a discussion where the arrow of time is what we want to explain. Here, I will therefore (in the context of the dynamics of the multiverse) discuss how this idea can be formulated in a time-symmetric way and how the asymmetric behavior of time may arise from this starting-point. The idea is to show that if we randomly choose one specific development of our multiverse, then the chance for this development to have an asymmetric behavior of time is enormously much larger than for a symmetric behavior. The underlying computations are certainly based on very crude approximations and simplifications. In Section 7, I will therefore finally list some of the objections that can be made.

In view of Price's criticism, there is a strong case for using a time symmetric terminology in a discussion of this kind. For instance, one could use a time-axis with directions "left" and "right" instead of "backward" and "forward". Nevertheless, the traditional way of thinking makes some parts easier to understand, so I have chosen not to be too persistent at this point.

## 2   The historical background

The cosmological perspective on the concept of entropy in a sense goes back to Boltzmann [6]. However, as a starting-point for the modern development we may take Gold's article [7] from 1962 (although this paper is in fact based on earlier results by Wheeler and Feynman, see [8]). Here, Gold argues that the second law of thermodynamics might be a consequence of an expanding universe. A natural consequence of this idea is that in a contracting universe, entropy would decrease and we are therefore naturally lead to consider a symmetric model for the universe with low entropy at both ends.

Today however, most cosmologists do not support the idea of such a symmetric behavior (see however Schulman [9] and also the discussion by the same author in [10]). Hence, most efforts since Gold have been concentrated on trying to understand why the behavior is asymmetric. There seem to have been at least three main strategies:

1. We can try to explain the arrow of time by making appeal to some kind of asymmetric property of the boundary conditions of the universe.
2. We can try to explain the arrow of time by making appeal to some kind of asymmetric property of the laws of physics themselves.
3. We can assume both the boundary conditions and the laws of physics to be essentially time-symmetric, but consider the arrow of time to be the result of some kind of broken symmetry.

The literature on this issue is very extensive. I will here only briefly recall some important ideas, partly to exemplify the above strategies but also because some of them are relevant for this paper. Good references for a more general discussion are Zeh [11] and Halliwell, Perez-Mercader, Zurek [10].

If we adopt the first strategy by assuming special boundary conditions, then the development of the entropy in between can essentially be understood starting from classical physics: if we somehow suppose that the state of the universe immediately after the Big Bang was exceptionally ordered, and also take for granted that the ultimate future of the universe (whether viewed as a Big Crunch or as an eternal expansion) will be very disordered in some sense, then the increase of entropy in-between appears quite natural.

It is also possible to consider closed models with essentially symmetric boundary conditions, but where the asymmetry arises from some unlikely event close to one of the ends. In this case there may still exist a definite arrow of time in between. However, this event would have to be so improbable that it is hard to accept it as a foundation for a theory of time without further motivation. It is part of the philosophy of this paper to reconsider this idea from the multiverse point of view,

Another way of looking at the problem has been suggested by Penrose: it could be that the growth of entropy is connected with the boundary conditions for the Weyl tensor [12]. This idea may be hard to confirm directly since the Weyl tensor is not readily observable. But it may still be a key to understanding the arrow of time, in particular in combination with other ideas from quantum cosmology,

An example of the second kind of strategy is given by the idea to connect the arrow of time with the time-asymmetric behavior of the K-meson. This idea appears to have been considered first by Sakharov [13] (for a somewhat more accessible account, see Davies [14]).

Examples of the third kind of strategy can be found in quantum cosmology. Often, the starting point is the Wheeler-deWitt equation (see [15] and also [11]). Hawking, in his early attempts to understand the arrow of time in terms of his "no boundary condition approach", adopted Gold's view. Later, after criticism from Page and Laflamme, he abandoned this idea and extended his analysis of the no boundary condition to conclude that there could be two types of behavior of the universe close to the end-points, one which fits nicely with a low-entropy Big Bang, and another one which fits with a high-entropy Big Crunch (see [16], [17]). However, the existence of these two different types of behavior does not in itself explain asymmetry. In fact, the main issue is to understand why the boundary behavior should be different at the two ends. In addition, it should be noted that parts of Hawking's reasoning may be somewhat controversial, e.g. the use of imaginary time.

There are also many other attempts to derive time-asymmetry from the Wheeler-deWitt equation within a more conventional framework (see Halliwell [18] and Zeh [11] for a discussion). One natural idea is to use the "superspace" formalism of Wheeler (see [19]) as an underlying framework for studying how different histories may interfere to yield asymmetry. This idea is also related to the approach in this paper.

It may also be said that any solution to the problem of time's arrow, based on the Wheeler-deWitt equation, would probably also in the end have to be an answer to the question what time itself is. A radical position has been taken by Rovelli ([20]) and Barbour ([21]): on the fundamental level, time may be a superfluous concept. In Barbour's setting, the only things which are actually real are different configurations in a timeless "Platonia". If we take this point of view, then the asymmetry of time amounts to a lack of symmetry in this timeless landscape, Again, this is a reasonable and interesting perspective. However, it does not in itself explain the asymmetry, so something additional has to be included.

In recent years, there has been an increasing interest in explanations of time asymmetry connected with cosmic inflation and the idea of multiple histories (see for example [22]). Such explanations may fit very well with various fundamental theories. However, there still seems to remain a considerable gap to bridge before this approach will be able to generate a convincing answer to the fundamental questions.

# 3   The multiverse

According to the standard interpretation, both classical Newtonian mechanics and quantum mechanics are deterministic theories: both Newton's equations and the Schrödinger equation can in principle be uniquely solved for all times once appropriate initial conditions have been specified.

In the Copenhagen interpretation of quantum mechanics, this determinism is in a sense destroyed by measurements where the outcomes are generally considered to be stochastic. According to the multiverse interpretation of quantum mechanics, initiated by Everett [23], each outcome of a measurement of a variable represents a real development. And the fact that a measurement has produced a certain value just indicates that what the observer thinks of as his universe is nothing more than a branch of the multiverse where the result of the measurement is the true value of the given variable. In this way, the multiverse interpretation can be said to restore determinism. It should be noted however, that determinism in this setting refers to the totality of all developments, the multiverse, and that for an observer confined to one branch of the multiverse, the outcome of an experiment still appears to be at random.

Nowadays, the idea of a multiverse has spread in different directions and acquired a number of different meanings (see Tegmark [24], [25]). In this paper however, I will only consider a very primitive and simplified multiverse. The multiverse will always be finite, not only in the sense that at each moment of time there will only be a finite number $N$ of particles, but in fact the total number of possible states for all particles will be finite. Also, I will only consider discrete time i.e. we split the time-axis up into short intervals of equal length which we take to be our unit of time.

The crucial property of the multiverse is that given a state at certain time $t$, there are in general several different developments leading to different states at time $t+1$. In theory, quantum mechanics could tell us how to compute the probability for different continuations. In the context of this paper however, we will usually not be able to consider the precise probability for each development but rather simply classify each transition from one state to another as "possible" or "impossible". Thus, for a given state at a given time $t$, there will be a certain number of states, at time $t+1$, which are "accessible" in the sense that a transition from the given state is possible. In this paper I will simplify the situation still further by assuming that the number $K$ (the "branching rate") of such accessible states is essentially independent of both the state and $t$. Thus, whether the universe is expanding or contracting, or whether the density is high or low within reasonable limits, we still assume that the number of choices for the development is fixed and is equal to $K$. A simple way of thinking about this is to say that at each step there is always a fixed number of choices open to each particle. Needless to say, this approximation may be a bad one in a realistic multiverse. But it could still be reasonable for the purpose of this paper.

It is important to note that we should *not* identify the states of the multiverse above with ordinary quantum mechanical states. In fact, different parallel universes should not be thought of as quantum mechanical states at all, but rather as a kind of "pseudo-states" which in this context represents a semi-classical approximation. Even if the underlying structure is quantum mechanical, the dynamics of these pseudo-states differs considerably from the time-development of ordinary quantum mechanical states: if different states correspond to different branches of the multiverse, time-development can not give isomorphisms between the sets of pseudo-states at different times, in spite of the fact that the underlying development of the wave-function is unitary. In fact, according to the multiverse perspective, it is generally assumed that all the different worlds which are parallel to ours originate from more or less one single state at the Big Bang.

Within this framework an enormous amount of different developments are possible:

**Definition 1** *A universe $U$ is a chain of states, one state $\Sigma_t$ at time $t$ for each $t$, with the property that the transition between $\Sigma_t$ and $\Sigma_{t'}$ is always possible according to the dynamical laws, where $t' = t + 1$.*

**Definition 2** *The multiverse $M$ is the set of all possible universes $U$ in the sense of Definition 1.*

Thus, we are actually studying a kind of superspace which is a huge graph where the states are nodes and where the edges connect states which are accessible from one another. When no ambiguity can arise, I will also sometimes refer to this superspace as the multiverse. It is implicit that "accessibility" is a symmetric relation, which physically corresponds to assuming the dynamic laws of physics to be symmetric with respect to time.

Within this multiverse, we also suppose that we can define an entropy function which somehow measures the amount of order/disorder of the possible states, and that the relation between the number $\Omega$ of states with entropy $S$ and the entropy $S$ itself is given by Boltzmann's famous formula ([26]):
$$S = k_B \log \Omega, \tag{1}$$
where $k_B$ is Boltzmann's constant.

Inverting this formula, we can also write
$$\Omega = W^S, \quad \text{where} \quad W = e^{1/k_B} \tag{2}$$
is a very large number (when using customary units). We conclude that at any time, the number of states $\Omega$ is an exponentially increasing function of the entropy $S$. Since we have assumed the number of states to be finite, it is clear that this can only hold for values of $S$ which are not too large. It is thus implicit in the following that the life-span of our multiverse is not long enough for the entropy to come close to being maximal, except (perhaps) near the end-points (i.e. near the Big Bang and a possible Big Crunch). In other words, states which have a reasonable chance to occur will in general be comparatively ordered. In fact, rough computations indicate (see [27] and [28]) that the entropy of our universe today is still rather low compared to what it could be, and with the present rate of increase it will continue to be so for a very long time still.

For reasons which will become clearer later, I will also assume that
$$0 \ll K \ll W, \tag{3}$$
where $K$ is the branching rate introduced in the beginning of the section. In fact, a multiverse where this is not fulfilled is unlikely to exhibit an arrow of time, see Section 4.

It is of course impossible to describe exactly how the entropy function $S$ develops with time. In the following I will make the very simple assumption that $S$ changes by the same amount during each unit interval of time, which partly can be said to amount to considering the entropy of our own universe as being a more or less linear function of time. Thus, after choosing appropriate units, we can say that if at a certain time $t$ the entropy takes the value $S$, then at time $t \pm 1$ it should be $S \pm 1$.

It is an important point for the following to note that in a multiverse where the dynamic laws are symmetric with respect to time, all such claims should also be reversible. If $S$ can grow by one unit per unit of time it can also decrease by one unit. So if there are $K$ states (with entropy $S + 1$) at time $t + 1$ accessible from a given state at time $t$, why should there not also be equally many states at time $t - 1$ (with entropy $S + 1$) which are accessible from the given state? This should indeed be the case, although in a realistic multiverse, it may of course not hold literally for each individual state, but rather it would be true on an average in a statistical sense. Thus, in the simplified multiverse of this paper I will in the following assume that starting from a given state, there are approximately $K$ accessible states in both directions of time, and whenever this simplifies the computations in a harmless way, I will actually assume that there are exactly $K$ such states. This symmetry will be further discussed in Section 6.

As a consequence, we see that among the $W^S$ states with entropy $S$ at a given time $t$, only $W^{S-1}K$ states can be accessible from states with lower entropy (i.e. entropy $S-1$) at time $t+1$, since there are only $W^{S-1}$ such states and only $K$ states with entropy $S$ can be accessed from each one of them, Therefore, if we randomly pick one state at time $t$ with entropy $S$, the chance that it should be possible to access from a state with lower entropy at time $t+1$ is

$$p = \frac{W^{S-1}K}{W^S} = \frac{K}{W} \ll 1, \tag{4}$$

in view of (3). We conclude that for a generic state at time $t$, the average number of accessible states with lower entropy at the next moment $t+1$ is $p \ll 1$, and the average number of accessible states with higher entropy at the next moment $t+1$ is $K - p \approx K \gg 1$. Obviously, the same argument applies also to the number of accessible states at time $t-1$.

## 4  The different arrows in the multiverse

Before we come to time asymmetry, it may be interesting to take a closer look at the arrow of time itself. It is customary in discussions of this kind to talk about *the* arrow of time, but there are actually several different arrows. According to the article by Hartle and Gellmann in [10], there are at least six or seven different arrows (see also Zeh [11]). Here, I will mainly deal with what I consider to be the two most important ones: *the thermodynamic arrow of time*, which is the concept based on entropy and *the psychological arrow of time* which is based on the observation that what is characteristic of the past is that it is unique and that we can actually remember it. All experience that we have seems to support the belief that the thermodynamic and psychological arrows are in some sense equivalent. But even if so, the equivalence is by no means a trivial one. And, especially in situations where we can not make use of our usual intuition, one should be very careful when identifying them.

It may here also be appropriate to mention a third arrow, namely *the cosmological arrow* which by definition points in the direction in which our universe expands. Thus, Gold's original suggestion that the entropy grows in the direction in which space expands, may be rephrased as an attempt to identify the cosmological arrow with the thermodynamic one.

To start with the thermodynamic arrow, the following formulation seems rather uncontroversial:

**Time's Arrow 1 (The increase of entropy)** *Whenever one macroscopic state $\Xi'$ succeeds another one $\Xi$ with respect to the development of time, the entropy of $\Xi'$ is larger than the entropy of $\Xi$.*

Suppose now that the multiverse starts from a low-entropy state ($S=0$) at the Big Bang. According to our assumptions about the dynamics of the multiverse in Section 3, we see that after one unit of time, we have $K$ possible developments corresponding to the possible accessible states. After two units of time we have $K^2$ possible developments, and after $t$ units of time we will have $K^t$ possible developments. How many of these have a monotonically increasing entropy? Again counting in the same way we see that we now at each step have on an average $K-p$ alternatives to choose between, so the number of strictly monotonic developments after $t$ units of time will be $(K-p)^t$ and hence the number of non-monotonic developments will be $K^t - (K-p)^t$. Thus we get (using (4)):

$$\frac{\text{"non-monotonic"}}{\text{"monotonic"}} = \frac{K^t - (K-p)^t}{(K-p)^t} = \left(1 + \frac{p}{K-p}\right)^t - 1 \approx \left(1 + \frac{1}{W}\right)^t - 1. \tag{5}$$

If $t$ is small compared to $W$, we get approximately (by Taylor expansion):

$$\frac{\text{"non-monotonic"}}{\text{"monotonic"}} \approx \frac{t}{W}. \tag{6}$$

Thus, if the life-span of our multiverse is small compared to $W$, a monotonic behavior of the entropy is much more common than a non-monotonic one. Without going into too many computational aspects of this theory, let me mention that for the kind of model I have in mind in Section 6, the constant $W$ will indeed be very large, so the requirement that $t/W$ should be small allows for a very long life-span of the multiverse.

The idea of the psychological arrow of time in the multiverse context can be formulated as follows:

**Time's Arrow 2 (The uniqueness of the past and non-uniqueness of the future.)** *Every state that we have ever observed seems to have a unique past but a non-unique future.*

We have in general a very good view of the historical development that has led us to the state that we observe now, but it is much harder to be sure about the consequences that the present state will lead to in the future. Stated somewhat differently, the past is the direction in which the development is uniquely traceable and there seems to be a unique chain of macroscopic states connecting our present reality back to the very early history of our universe. As for the future on the other hand, in any physical situation there seem to be phenomena which are genuinely unpredictable like, for instance, radioactive decay. According to the multiverse point of view, every such event represents a fork in the road towards the future.

So why should the history be unique but not the future? If we as in (3) assume that $K \ll W$, then at a given time $t$ (i.e. $t$ units of time after the Big Bang), the set of states resulting from monotonically increasing developments (i.e. with entropy $S = t$, according to the choice of units in Section 3) will be enormously much larger than the set of parallel universes up to time $t$, since the former number is $W^t$ and the latter is given by $K^t$. Consequently, if we assume that we can argue probabilistically, the chance that for a randomly chosen state there will be a development leading back to the original state at the Big Bang is very small. And the chance that there will be two or more such developments is even much smaller. Thus, an observer in almost any of the possible universes will see only one possible history. In other words, he will observe a psychological arrow in the sense of the above definition.

In this sense, both arrows can be said to be probabilistic consequences of the dynamics in the simple multiverse of Section 3. It is interesting however, to observe that this does not mean that they are entirely equivalent. Even if they both appear and point in the same direction under reasonable assumptions, the circumstances under which they cease to exist differ considerably: The thermodynamic arrow appears to be an essentially classical concept, and it will persist as long as formula (1) is applicable, i.e. as long as the universe is still sufficiently ordered (at least if the life-span of the multiverse is not too long).

The psychological arrow on the other hand is inherently quantum mechanical. It will cease to exist when the number of possible developments up to a certain time $t$ will approach the total number of states at that time. When this happens will depend on the size of $K$ (which is related to Planck's constant $h$) and has little to do with the entropy.

In this paper, I will not say much more about the relationship between the arrows. I will simply assume it to be a fact that, in a simple multiverse of the kind discussed in Section 6, every development is unique in the direction of decreasing entropy but highly non-unique in the direction of increasing entropy.

# 5   What is entropy?

There are many different aspects of the concept of entropy and there are also many different definitions of entropy which do not appear to be fully compatible (see [29]). Nevertheless, for the purpose of this paper, I shall take the point of view that as long as we study stationary (or quasi-stationary) problems in bounded regions of fixed volume, entropy is an essentially well understood concept. Also, clearly there may be problems with the concept of entropy connected with black holes, dark energy and similar phenomena, but these will be left out of the discussion in this paper. However, there is one problem which we can not avoid when discussing time's arrow, and where our theoretical understanding is much more uncertain and empirical evidence is lacking. This is the question raised by Gold [7]: Does the expansion of the multiverse in itself contribute to the growth of entropy? Partly, this is of course a question about definitions. In particular, we have to be able to tell how many states there are and how to count them.

According to the philosophy of this paper (see Section 1), I will adopt an essentially classical perspective on this question. In fact, at this point I would find it quite satisfactory if I could construct a theory for a multiverse similar to some kind of lattice gas (see e.g. [30]). On the other hand, there seems to be no particular reason for not as well considering the somewhat more realistic case of an ordinary space-time split up into microscopic elementary cells with approximately equal volume, where we consider the state of the multiverse to be determined by specifying which cells are occupied by which particles. Or for that part, to consider the same kind of splitting in a classical phase-space with suitable restrictions on the total energy.

Clearly, the number of states of a single particle in such a multiverse will be proportional to the volume of space at the given time, and the number of possible states of $N$ particles will be approximately proportional to $V(t)^N$ where $V(t)$ is the volume of space at time $t$. (if we assume the number of possible one-particle states to be large compared $N$). Thus, near the Big Bang or the Big Crunch, the number of possible parallel worlds may be comparatively small, but in between it can grow to something enormously large.

If we consider the simplest case where all the particles are independent, then taking logarithms and using (1), we see that the entropy as a function of the total volume of space at a time $t$ can be expressed as
$$S = Nk_B \log V(t). \tag{7}$$
In fact, this is exactly the formula for the entropy of a classical ideal gas (see [4]) when temperature is kept constant. In classical physics, this formula can be interpreted as saying that by expanding, the gas looses a part of its ability to do useful work on the exterior (hence entropy grows). In the cosmological context however, such an interpretation becomes meaningless: the expansion and contraction of space is simply an outer requirement which the gas must adapt to. Nevertheless, formula (7) makes perfectly good sense, and moreover, Gold's view also makes good sense: If we imagine a multiverse consisting of such an ideal classical gas (which is in a state of maximal entropy) shortly after the Big Bang, then we see that if first space expands, the entropy will grow. And if afterwards space contracts down to its previous size, the entropy will decrease to its original value.

On the other hand, this behavior seems to have very little to do with the second law of thermodynamics as a manifestation of growing disorder: the gas is all the time in a state of maximal entropy and after having first expanded from a given size and then again contracted down to the original size, the gas is simply back in the same macroscopic state.

In this cosmological example, we are faced with a choice: in a dynamical space-time, we must either give up the classical definition of entropy (based on formula (1)), or we must give up our interpretation of the second law as a manifestation of growing disorder. My conclusion is that it is preferable to modify formula (1). However, such a modification is a non-trivial and perhaps painful step. Therefore, I suggest the following two conditions which such a modification should satisfy:

- Our definition of dynamic entropy in a dynamical space-time must coincide with our usual definition in the case of a stationary bounded region of fixed volume.
- Our definition of dynamic entropy should be scale-invariant in the sense that if all distances in space are scaled by the same factor, the entropy remains unchanged.

Thus for instance, in the example with the ideal gas above, the dynamic entropy should be constant throughout the whole process of expanding and contracting. However, an attempt to introduce such a modified entropy in more general situations raises many question which are not easy to answer. One possible way to try to understand the behavior of entropy, could be to split the entropy up into two parts:

$$S_{total} = S_{geometric} + S_{dynamic}, \tag{8}$$

where $S_{geometric}$ is so to speak the trivial part caused by the expansion of the universe, and $S_{dynamic}$ is the part which is actually connected with the growth of disorder. It is not obvious that this splitting makes good sense in a realistic multiverse, since the expansion of space is coupled to various other processes involving entropy. But it does seem reasonable for instance in a multiverse where the interactions are rather weak. I will in the following use formula (8), with $S_{geometric} = Nk_B \log V$ as in the case of an ideal gas, as a coarse approximation to the truth.

Summing up, when we speak of the second law of thermodynamics in this paper, this will refer to the following statement:

**Principle 1** *The dynamic entropy of our universe is non-decreasing.*

If we now in formula (8) divide by $k_B$ and exponentiate, and then combine with (2) and (7) above, we get the following formula for the number of states with entropy $S$:

$$\Omega \propto V(t)^N \times W^S, \tag{9}$$

where $S$ now refers to the dynamic entropy. From now on, I will simply drop the word "dynamic" and just speak of entropy when no confusion is possible.

**Remark 1** *It should be noted that our universe is quite far from being an ideal gas, and the expansion of the universe does not take place at constant temperature. Nevertheless, for the purpose of this paper, I shall consider formula (9) to be good enough as an estimate of the number of possible "states of parallel worlds" with entropy $S$ at time $t$.*

# 6 The asymmetry of time

In this section we will consider the main issue of this paper: Why is time asymmetric? The answer that I propose is that if we consider the multiverse to be a huge probability space and apply certain physical assumptions to assign to each universe a probability weight, then it may turn out that developments with low entropy at one end and high entropy at the other are by far the most probable ones. The underlying laws of physics are completely symmetric and so is the result: the probability for low entropy close to one endpoint and high entropy close the other equals the probability for the reversed behavior. However, for an observer confined to a specific universe, the arrow of time will appear to have a definite direction.

Rephrasing this idea in still another way, we may say that (according to this model), among all observers in universes parallel to ours, $50\%$ will perceive our Big Bang as the Big Crunch, and $50\%$ will perceive it the same way as we do.

For definiteness and simplicity, I will consider a closed Friedmann space-time ([31], [32] and [11]). In this setting, let $[-T_0, T_0]$ be the time-interval from the Big Bang to the Big Crunch. It will be convenient to split the life-span of the multiverse into three different phases: Thus, we choose symmetric moments of time $-T_1$ and $T_1$ close to the end-points of the interval $[-T_0, T_0]$ so that we can write

$$[-T_0, T_0] = [-T_0, -T_1] \cup ]-T_1, T_1[ \cup [T_1, T_0]. \tag{10}$$

Coarsely speaking, this may be thought of as a kind of idealized division into "the extreme initial phase", "the normal phase" and "the extreme final phase" of the multiverse. I have no definite opinion about what the length of the extreme phases should be, but it seems reasonable to assume them to be very short (much shorter than a second). I will also assume that at times $-T_0$ and $T_0$, every universe will be in a unique state with entropy zero.

During the normal phase, usual physics more or less as we know it applies. But during the extreme phases the world that we are trying to model may be very different from what we are used to and the arrow of time may not have any meaning. Clearly, the transition from one phase to another should in a more realistic model be treated as gradual. However, to illustrate the general mechanisms of this paper, this approximation seems appropriate.

The following is an extension of Boltzmann's idea ([6]) about the second law as a manifestation of the fact that the universe passes (as time evolves) from less probable states to more probable ones. The formulation below is essentially a summation of the discussion at the end of Section 3, and the reader may note that is manifestly time symmetric:

**Statistical Assumption 1** *Given any state of a universe during the normal phase (with entropy far from being maximal), the number $K$ of accessible states with higher entropy (in both directions of time) is enormously large, and the number of accessible states with lower entropy (in both directions of time) is very small (or rather, the probability $p$ for finding any such accessible state is very small). As before, we will consider $K$ and $p$ to be essentially independent of time and of the particular state.*

**Remark 2** *It should perhaps be emphasized that the assumption about complete symmetry between developments in both directions of time at each specific moment of time is not very realistic in a universe like ours. In fact, many processes (like for instance the emission of light from a source) tend to so to speak preserve a memory of the direction of time for a long time.*

*It could be that the assumption is rather reasonable in, say, a gas where the average time between interactions of a particle is short compared to the unit of time. Such a gas however, is of course quite far from being a good model for our universe.*

*My conclusion is that if we could explain the asymmetry of time in such a gas, it would still be a big step forward. Alternatively, we may choose to perceive the above assumption as only being true in a very general average sense.*

To make things as simple as possible, let us not consider the full multiverse, but rather concentrate on the subset consisting of universes where the arrow of time switches direction only once. Of course, it is quite possible to imagine a universe where the direction of the arrow switches several times or where there is no arrow at all (the entropy may for instance be constant). Nevertheless, according to the discussion in Section 4, in the framework I have chosen, a monotonic behavior of the arrow is much more likely than a non-monotonic one, so for the purpose of contrasting a symmetric behavior (Gold's view) with a non-symmetric one, the present setup in view of its simplicity seems adequate.

Summing up, we will below only consider universes of the following kind: At time $-T_0$ the entropy is zero (complete order). During the first extreme phase, the entropy will grow from zero to some value $x$. After this, during the normal phase from $-T_1$ to $T_1$, the entropy will first grow from this value $x$ up to a maximal value $S_m$ at some time $t_m$, and then it will decrease down to some other value $y$ at $T_1$, after which it then finally shrinks down to zero again during the final extreme phase. It is important to note that $t_m$ may very well coincide with $-T_1$ or $T_1$; this is actually as we shall see the most probable case. We also note that (according to the very simple dynamical assumptions about the entropy in Section 3, stating that it can only grow or decrease with one unit per unit of time) given $x$ and $y$, the point $t_m$ is completely determined by the formula

$$t_m = \frac{y-x}{2}, \tag{11}$$

since the graph of the entropy function consists of two linear segments with slope plus and minus one. A short computation shows that the maximal value of the entropy will be given by

$$S_m = S(t_m) = T_1 + \frac{x+y}{2}, \tag{12}$$

To each universe we can now assign a certain probability weight $\omega$ which factors into three parts:

$$\omega = \omega_{-1}\omega_0\omega_1. \tag{13}$$

Here $\omega_{-1}$ and $\omega_1$ represent the weights for the development from $-T_0$ to $-T_1$ and from $T_1$ to $T_0$ respectively, whereas $\omega_0$ refers to the normal phase in between.

During the extreme phases, the multiverse is extremely small and quantum effects dominate. Starting from the completely ordered states at $-T_0$ and $T_0$, almost anything can happen with positive probability. On the other hand, the extreme phases have a very short duration, so we will assume that the most probable development is that very little happens at all, i.e. the universe enters the normal phase in a very ordered state. In any case, whatever happens is a very complicated process which depends on an very large number of more or less independent phenomena. Needless to say, there are many developments from $-T_0$ to $-T_1$. But since the entropy is really the only variable which interests us, I will simplify the situation somewhat by just considering one state at $-T_1$ for each (integer-)value of the entropy (and similarly at $T_1$). This is of course a simplification, but as long as the number of developments during the extreme phases is small compared to corresponding number of developments during the normal phase, it appears to be a rather harmless one. From a statistical point of view, it is now very natural to assume that the probabilities for the possible values $x$ and $y$ of the entropy at times $-T_1$ and $T_1$ are given by a Poisson distribution:

$$\omega_{-1} \sim \frac{\lambda^x}{x!}, \qquad \omega_1 \sim \frac{\lambda^y}{y!}, \tag{14}$$

where $\lambda$ is some positive number. If we assume $\lambda \ll T_1$ to be moderate in size, only rather small values of $x$ and $y$ will give probability weights which are not extremely small. On the other hand, it is important to observe that also large values of $x$ and $y$ have a non-zero weight. As has previously been said in Section 3, we will during the normal phase only distinguish between possible and impossible developments. In this context this amounts to only considering the possible values $0$ and $1$ for $\omega_0$.

With this set-up, we can now define a two-variable function $f(x,y)$ which is the sum over all probabilities of universes where the entropy takes the values $x$ and $y$ at times $-T_1$ and $T_1$ respectively. According to (13) and (14), we have

$$f(x,y) = N(x,y) \frac{\lambda^x}{x!} \frac{\lambda^y}{y!}, \tag{15}$$

where $N(x,y)$ denotes the number of possible universes with entropy $x$ at time $-T_1$ and entropy $y$ at time $T_1$. $f(x,y)$ thus gives the (un-normalized) probability for a randomly chosen universe to have entropy $x$ at time $-T_1$ and $y$ at time $T_1$.

The proposed explanation of time-asymmetry is now contained in the following:

**Claim 1** *Under appropriate circumstances (i.e. appropriate choices of the parameters in the model), $f(x,y)$ takes very sharp global maxima when $x \approx 0$ and $y \approx 2T_1$ (corresponding to a monotonically increasing entropy), and when $x \approx 2T_1$ and $y \approx 0$ (corresponding to a monotonically decreasing entropy).*

The first observation to this end is that a global maximum of $f(x,y)$ must be attained when $x \approx 0$ or $y \approx 0$. This is simply because $N(x+s, y+s)$ is a decreasing function of $s$ (this follows from (16) below, since $G(t_m)$ according to (9) increases with $S_m$), and the factors $p(x) = \lambda^x/x!$ and $p(y) = \lambda^y/y!$ in (14) will also be decreasing functions except for very small $x$ and $y$. Since the function $N(x,y)$ will (due to the assumption that the entropy can only grow by one unit per each unit of time) only be non-zero when $|x - y| \leq 2T_1$, it follows that the maximum of $f(x,y)$) can be attained only inside the square $\Delta = [0, 2T_1] \times [0, 2T_1]$.

To compute $N(x,y)$, we will again make use of a very general statistical idea: It is easy to compute the number of possible developments from time $-T_1$ up to the point $t_m$ where the entropy is maximal. Using the multiplicative counting as in the discussion leading to (5) we get $K^{T_1+t_m}$ such developments. Similarly, we may compute the number of backwards development from $T_1$ back to $t_m$, giving $K^{T_1-t_m}$. So the question is really how many of the forward developments from $-T_1$ which fit onto backward developments from $T_1$. To estimate this we will simply assume that the chances for a state to be accessible by developments from $-T_1$ and $T_1$ respectively, can be treated as statistically independent variables.

**Claim 2** *Suppose that $K$ is such that the chance for two different developments starting from some states at time $-T_1$ to meet before time $T_1$ is negligible small. Then the number $N(x,y)$ of universes such that $S$ is maximal at $t_m \in ]-T_1, T_1[$ is proportional to*

$$\frac{1}{G(t_m)}, \tag{16}$$

where $G(t_m)$ is the number of states of the global geometry at time $t_m$ with entropy $S_m$.

In fact, the number of such universes is given by the total number $K^{T_1+t_m}$ of developments with monotonically growing entropy from $-T_0$ up to $t_m$, multiplied by the probability that such a development will fit onto a development with monotonically decreasing entropy from $t_m$ to $T_0$. From our heuristic assumption about statistical independence, we see that the latter is given by the number $K^{T_1-t_m}$ of such possible developments divided by $G(t_m)$. Thus

$$N(x,y) \propto K^{T_1+t_m} \times \frac{K^{(T_1-t_m)}}{G(t_m)} = \frac{K^{2T_1}}{G(t_m)} \propto \frac{1}{G(t_m)}. \tag{17}$$

According to formula (9) the total number of states $G(t_m)$ can be expressed as

$$G(t_m) = (V(t_m))^N \times W^{S_m} \tag{18}$$

where $V(t_m)$ is the volume of space at time $t_m$, which again is proportional to $R(t_m)^3$ where $R(t)$ in a closed

Friedmann space-time is given by a cycloidal function $R(t)$ on $[-T_0, T_0]$ (see [32]). Summing up, we arrive at (up to a multiplicative constant) the following expression for $f(x,y)$ (using (11), (12), (15) and (16)):

$$f(x,y) = R(\tfrac{y-x}{2})^{-3N} \times W^{-(x+y)/2} \times \frac{\lambda^{x+y}}{x!y!}, \qquad (19)$$

Although the cycloid function $R(t)$ is not readily expressible by elementary functions, it is now possible to analyze the function $f(x,y)$. It may in fact be somewhat more natural to investigate the logarithm $g(x,y) = \log f(x,y)$:

$$g(x,y) = -3N \log R(\tfrac{y-x}{2}) - \tfrac{x+y}{2}\log W + \lambda(x+y) - x\log x + x - y\log y + y, \qquad (20)$$

where I have also used Stirling's formula $\log(x!) \approx x\log x - x$.

It is not true that the function $f(x,y)$ (or equivalently $g(x,y)$) will achieve its maximum at the points indicated in Claim 1 for all possible values of $N$, $T_1$, $\lambda$ and $W$. What *is* true however, is that this is what happens for the first term $-3N\log R(\tfrac{y-x}{2})$ in (20), since the function $-\log R(t)$ obviously takes its maximum when $t = \pm T_1$, and with $t = \tfrac{y-x}{2}$ this corresponds to the two points $(0, 2T_1)$ and $(2T_1, 0)$ in the square $\Delta$. If the other terms in (20) are sufficiently large, this need not be true for $g(x,y)$, but if on the other hand they are sufficiently small it will be. Coarsely speaking, if $N \gg T \times W$ there will indeed be an asymmetric behavior, where $T$ is the total life-span of the multiverse (compare the Mathematica plot to the right in figure 1), but if on the other hand $N \ll T \times W$ the result will instead be a Gold's type of multiverse with low entropy at both ends (to the left in figure 1).

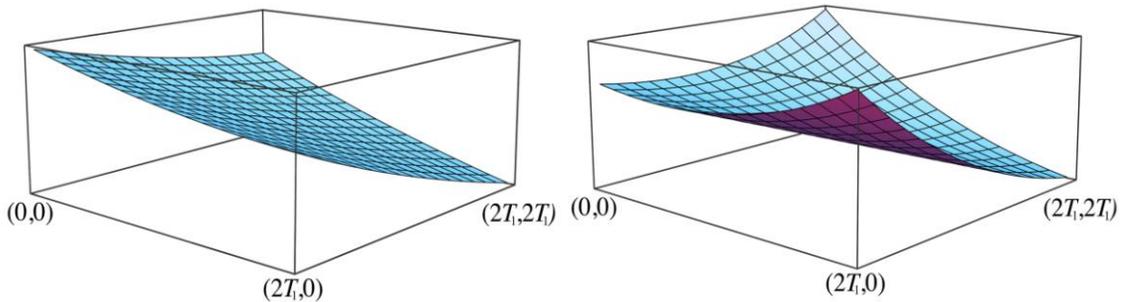

Figure 1 These plots show $g(x,y)$ as a function of the entropies at times $-T_1$ and $T_1$. Thus the value at $(0,0)$ gives the (logarithm of) the probability for a universe with low entropy at both ends, whereas the value at e.g. $(2T_1, 0)$ gives the probability for low entropy at one end and high entropy at the other. In the plot to the left we have a Gold's type of situation where the final terms in $g(x,y)$ dominate. To the right we have the situation where the first term dominates and the probability is largest at $(2T_1, 0)$ and $(0, 2T_1)$. Thus in this case, the behavior of time is asymmetric.

**Remark 3** *It is clear that the very coarse model in this Section should not seriously be interpreted in a quantitative way. Still it may be interesting to ask what kind of function $f(x,y)$ would give reasonable estimates of $N$, $W$ and $T$?*

*First, let us simply put $N = 10^{80}$ which correspond to a standard estimate of the number of baryons in the observable universe. $W$ and $S$ are related as in (2), and if we make use of the fairly recent estimates of the total entropy of our universe in [28], it appears that if we just consider ordinary matter mainly concentrated in stars and the corresponding stelar entropy then it is possible that $f(x,y)$ could exhibit the right kind of behavior, but it is not clear to me how to evaluate various uncertain factors. If we however take into account the total entropy according to the quoted paper, then we get an entropy which is larger by a factor $10^{24}$, which apparently makes the second term in (20) dominate over the first. Again, it is not clear to me what conclusion should be drawn even if we disregard all the numerous defects of the model: this entropy mainly steams from super-massive black holes and also from radiation, both of which are not included here.*

# 7  Discussion and comments

Let me finally list some of the (very many) objections which can be made to this model

- Simplifications are abundant. To mention a few: The constancy of $K$ and $W$, the abrupt change of the direction of times arrow at $t_m$, the assumptions about the extremely simplified behavior of the entropy, the sharp division into phases with different dynamic behavior (extreme phases and the normal phase), the exclusion of major sources of entropy etc.

- The model is based on macroscopic assumptions that may simply be wrong. We do not know that we live in a closed Friedmann space-time, and indeed many cosmologists prefer open models, to a large extent due to the discovery of the presently accelerating expansion of our universe. Also, the dynamic entropy of Section 5 may be controversial, and besides most cosmologists believe that the dominating part of the entropy is connected with phenomena which I have completely neglected in this paper, like black holes, radiation, dark energy etc.

- The treatment can not be said to be quantum mechanical.

Needless to say, it would be desirable to improve on all these defects, and I will welcome every attempt to do so. On the other hand, it is not clear how far it is possible to extend this kind of attack, due to the enormous complexity of the multiverse. A possible alternative approach to the kind of explanation of time's arrow which I have suggested in this paper would be to use computational combinatorics and graph theory to study small multiverses governed by the kind of semi-classical dynamics discussed in Section 3. This certainly leads to difficult but perhaps not impossible problems.

Martin Tamm,
Dept of Mathematics,
Univ. of Stockholm,
S-106 91 Stockholm,
Sweden.
matamm@math.su.se